\documentclass[aps,prd,onecolumn,groupedaddress,nofootinbib]{revtex4}         

\usepackage[english]{babel}

\usepackage[letterpaper,top=2cm,bottom=2cm,left=3cm,right=3cm,marginparwidth=1.75cm]{geometry}

\usepackage{amsmath}
\usepackage{graphicx}
\usepackage[colorlinks=true, allcolors=blue]{hyperref}
\usepackage{xcolor}

\begin{document}

 \title{Is Cosmic Birefringence model-dependent?}

\author{Lu Yin$^{1}$}
\email{lu.yin@apctp.org}
\author{Joby Kochappan$^{1}$}
\author{Tuhin Ghosh$^{2}$}

\author{Bum-Hoon Lee$^{3, 4}$}

\affiliation{$^{1}$Asia Pacific Center for Theoretical Physics, Pohang, 37673, Korea}
\affiliation{$^{2}$National Institute of Science Education and Research, An OCC of Homi Bhabha National Institute, Bhubaneswar 752050, Odisha, India}
\affiliation{$^{3}$Center for Quantum Spacetime, Sogang University, Seoul 121-742, South Korea}
\affiliation{$^{4}$Department of Physics, Sogang University, Seoul 121-742, South Korea}

\begin{abstract}
Exciting clues to isotropic cosmic birefringence have recently been detected in the $EB$ cross-power spectra of the polarization data of the cosmic microwave background (CMB). Early Dark Energy (EDE) models with a pseudoscalar field coupled to photons via a Chern-Simons term can be used to explain this phenomenon, and can also potentially be used to simultaneously resolve the $H_0$ tension.
In this work we incorporate an early dark energy scalar field, including a Chern-Simons coupling, into an existing Boltzmann solver and numerically recover the $EB$ cross-power spectrum for two models in the literature; the $\alpha$-attractor, and the Rock `n' Roll field. We find that both the models fit the $EB$ spectra, and the $EB$ spectra alone do not possess sufficient constraining power to distinguish the two models based on current data.
\end{abstract}

\maketitle

\section{Introduction}

The observation of the Cosmic Microwave Background (CMB) is one of the most important discoveries in 20th century physics, which helped to establish the standard $\Lambda$CDM cosmological model \cite{Komatsu:2014ioa, Planck:2018vyg, SPT-3G:2021eoc, BICEP:2021xfz, SPIDER:2021ncy}.
However, with recent and more precise observations, the CMB in conjunction with other data suggest that there may be new physics beyond the standard model\cite{Abdalla:2022yfr}. Here we consider two of the most compelling issues;  ``cosmic birefringence"
\cite{Komatsu:2022nvu} and the ``Hubble tension" \cite{Bernal:2016gxb}.

Cosmic birefringence is the rotation of the plane of linear polarization of photons \cite{Carroll:1989vb, Carroll:1991zs, Harari:1992ea}.
Recent measurements indicate that the Planck polarization data exhibit a  3.6$\sigma$ 
signal of a non-zero cosmic birefringence angle, $\beta={0.342^{\circ}}^{+0.094^\circ}_{-0.091^\circ}  (68\% C.L.)$  \cite{Minami:2020odp,Eskilt:2022cff}. This new phenomenon comes from the cross-correlations between the $E$ and $B$ mode in CMB polarization \cite{Lue:1998mq}.

The $E$ and $B$ modes of the CMB transform under parity according to $E_{\ell m} \to (-1)^{\ell} E_{\ell m}$ and $B_{\ell m} \to (-1)^{\ell+1}B_{\ell m}$, and the auto power spectra of these quantities are invariant under the transformation. However, the cross power spectrum changes sign and so non-zero values of $C_{\ell}^{EB}$ could indicate parity violation; new physics beyond the standard model of cosmology, such as axion-like pseudoscalar fields that couple to the electromagnetic tensor and its dual as $g \phi F_{\mu\nu} \tilde{F}^{\mu\nu}$ \cite{Nakatsuka:2022epj, Murai:2022zur}. 
Such a term in the Lagrangian would rotate the plane of polarization of the CMB photons and generate a non-zero $EB$ cross power spectrum. 
Additionally, such a term can convert some power from the $E$ to the $B$ modes. Data from upcoming experiments like the Simons Observatory \cite{SimonsObservatory:2018koc}, ground based CMB-S4 \cite{Abazajian:2019eic} and LiteBIRD \cite{LiteBIRD:2022cnt} could potentially detect this signal in the CMB power spectra. LiteBIRD is expected to have the sensitivity to detect both the signal in the $EB$ spectra, as well as the induced $BB$ components.

In parallel to these developments, there is now robust evidence of a $\sim 4-5\sigma$ discrepancy between the Hubble parameter inferred using cosmological scale data such as the CMB and BAO, and directly measured values via the distance ladder \cite{Bernal:2016gxb, Riess:2021jrx}. This challenge to the standard model is yet to be resolved, with possible explanations including unknown systematics in the distance ladder, non-standard dark energy models and some modification to the expansion history at high redshift. This latter possibility has gained traction as one of the few ideas that can ameliorate the tension in a way that does not introduce new tensions between cosmological parameters. An early dark energy (EDE) component, such as an axion-like field, can potentially explain both the Hubble tension and generate a small signal in the CMB $EB$ cross power spectrum. 

Multiple early dark energy models have been proposed in the literature \cite{Caldwell:2003vp, Smith:2019ihp,Berghaus:2019cls, Alexander:2019rsc, Chudaykin:2020acu, Agrawal:2019lmo, Niedermann:2019olb, Freese:2004vs, Ye:2020btb, Akarsu:2019hmw, Lin:2019qug, Yin:2020dwl, Braglia:2020bym}. Especially, early studies introduced a string-like axion with potential $V(\phi) \sim (1 - \cos[\phi/f])^{n}$~\cite{Smith:2019ihp},  where $f$ is the axion decay constant, with phenomenological parameter $n > 1$ required to ensure that the energy density of the field decays sufficiently quickly in the late Universe. Other models include scalar fields inspired by inflationary $\alpha$-attractors~\cite{Braglia:2020bym}, which are a class of models that predict spectral index $n_{s}$ and tensor-to-scalar ratio $r$ values that are practically independent of the specific functional form of the potential $V(\phi)$, and Rock `n' Roll model with power law potentials $V \propto (\frac{\phi}{M_{pl}})^{2n}$ \cite{Agrawal:2019lmo}. All of these models function in similar ways -- introducing a non-negligible dark energy component in the early Universe which modifies the sound horizon at decoupling $r_{s}$, followed by a rapid decay of energy density. 

In this work, we consider the phenomenology of these different models when an explicit coupling between scalar field and electromagnetic field tensor is included. In particular, we search for potential signals in the $EB$ cross power spectrum that might be used to distinguish different early dark energy models. We consider how various models modify the Boltzmann equations and implement them into numerical solvers, extracting the power spectra and considering how their shape and amplitudes can be used to discriminate between competing theories.  In particular both the $EE$ and $BB$ modes will be modified by Cosmic Birefringence, so we compare the power spectra from different models. 

The paper will proceed as follows. In Section \ref{sec:2} we review the Boltzmann equations and how they are modified by the presence of an explicit coupling between a scalar field and photons. In Section \ref{sec:3} we review the early dark energy models that we explore. Our main results are presented in Section \ref{sec:4}; the $EB$ power spectra and the tools we use to discriminate between different models. We conclude in Section \ref{sec:5}.

\section{Boltzmann Equation For Cosmic Birefringence}
\label{sec:2}

Through the Chern-Simons term, the Lagrangian density with coupling between pseudo-scalar field and photons can be written as \cite{Murai:2022zur}
\begin{equation}
    \mathcal{L}=-\frac{1}{2}\left(\partial_\mu \phi\right)^2-V(\phi)-\frac{1}{4} F_{\mu \nu} F^{\mu \nu}-\frac{1}{4} g \phi F_{\mu \nu} \tilde{F}^{\mu \nu},
\end{equation}
where the pseudo-scalar field $\phi$ has a canonical kinetic term and potential $V(\phi)$, $g$ is the Chern-Simons coupling constant of mass dimension -1, and $F_{\mu\nu}$ and $\tilde{F}_{\mu\nu}$ correspond to photon field's strength tensor and its dual, respectively.

Due to the Chern-Simons term in the Lagrangian, the photon dispersion relation is modified as \cite{Carroll:1989vb, Carroll:1991zs,Harari:1992ea}
\begin{equation}
\omega_{\pm} \simeq k \mp \frac{g}{2}\left(\frac{\partial \phi}{\partial t}+\frac{\textbf{k}}{k} \cdot \nabla \phi\right)=k \mp \frac{g}{2} \frac{\mathrm{d} \phi}{\mathrm{d} t}
\end{equation}
where $+$ and $-$ are the right- and left-hand circular polarization states of photons, $\omega_{\pm}$ correspond to the $\pm$ helicity states' angular frequency. We adopt the convention of the right-hand coordinate of the $z-$axis as the direction of photon propagation. 

The rotation of the plane of linear polarization depends on the dispersion relation of the helicity. We can use the fact that $\omega_\pm$ is much larger than the time evolution of $\phi$ in the WKB limit. Using the WKB approximation, we can give the angle of rotation from a given time $t$ to the current time $t_0$ by
\begin{equation}
\label{beta}
\beta(t)=-\frac{1}{2} \int_t^{t_0} \mathrm{~d} \tilde{t}\left(\omega_{+}-\omega_{-}\right)=\frac{g}{2}\left[\phi\left(t_0\right)-\phi(t)\right] .
\end{equation}

\noindent With our choice of conventions, $\beta > 0$ means that the linear polarization rotates clockwise in the sky, and the $z$-axis is in the direction of the observer's line of sight. We follow the linear polarization position angle conventions used in \cite{Komatsu:2022nvu}.

Since the early dark energy is {{modeled}} using a pseudo-scalar field.
In order to describe the  evolution of CMB polarization including the cosmic birefringence effect, we use the modified Boltzmann equation with contribution from rotation angle $\beta$ -- 

\begin{equation}
{ }_{\pm 2} \Delta_P^{\prime}+i q \mu_{\pm 2} \Delta_P=  \tau^{\prime}\left[-{ }_{\pm 2} \Delta_P+\sqrt{\frac{6 \pi}{5}}{ }_{\pm 2} Y_2^0(\mu) \Pi(\eta, q)\right]  \pm 2 i \beta^{\prime}{ }_{\pm 2} \Delta_P,
\end{equation}
where $\eta$ is the conformal time. In Fourier space, the wave vector is $q$, and the angle between $q$ and photon momentum is expressed as $\mu$. In this equation, ${ }_{\pm 2} Y_\ell^m$ denote the spin-2 spherical harmonics, and the polarization source term is $\Pi(\eta, q)$.  The function ${ }_{\pm 2} \Delta_P$ is the Fourier transform of $Q\pm i U$, where the $Q$ and $U$ are the real and imaginary parts of the Stokes parameters of linear polarization, respectively.

To express the Boltzmann equation in terms of ${ }_{\pm 2} Y_2^0(\mu)$, we can expand ${ }_{\pm 2} \Delta_P$ as 
\begin{equation}
{ }_{\pm 2} \Delta_P(\eta, q, \mu) \equiv \sum_\ell i^{-\ell} \sqrt{4 \pi(2 l+1)}{ }_{\pm 2} \Delta_{P, l}(\eta, q)_{\pm 2} Y_\ell^0(\mu).
\end{equation}
In general, $\beta$ evolves with conformal time and we obtain the solution of the Boltzmann equation with rotation angle $\beta(\eta)$ as 
\begin{equation}
\begin{array}{rl}
{ }_{\pm 2} \Delta_{P, l}\left(\eta_0, q\right)=-\frac{3}{4} \sqrt{\frac{(l+2) !}{(l-2) !}} \int_0^{\eta_0} \mathrm{~d} \eta \tau^{\prime} e^{-\tau(\eta)} \Pi(\eta, q) \times \frac{j_\ell(x)}{x^2} e^{\pm 2 i \beta(\eta)} , 
\end{array}
\label{Boltzmann}
\end{equation}
where $x=q(\eta_0-\eta)$, and $j_\ell(x)$ is the spherical Bessel function.

The CMB polarization power spectra can be written as
\begin{equation}
C_\ell^{X Y}=4 \pi \int \mathrm{d}(\ln q) \mathcal{P}_s(q) \Delta_{X, l}(q) \Delta_{Y, l}(q) ,
\end{equation}
where $\mathcal{P}_s$ corresponds to the primordial scalar curvature power spectrum, $X, Y$ are labels for the $E$ or $B$ mode CMB polarization, which can be directly inferred from Eq.\ref{Boltzmann} as
$\Delta_{E, l}(q) \pm i \Delta_{B, l}(q) \equiv-{}_{\pm 2}\Delta_{P, l}\left(\eta_0, q\right)$.
If $\beta$ is constant, we can rewrite the $E$ and $B$ fluctuations as 
\begin{equation}
\Delta_{E, \ell} \pm i \Delta_{B, \ell}=e^{\pm 2 i \beta}\left(\tilde{\Delta}_{E, \ell} \pm i \tilde{\Delta}_{B, \ell}\right), 
\end{equation}
where $\tilde{\Delta}$ are the $E$ and $B$ fluctuations in the absence of cosmic birefringence. 
Here, parity violation is present in the $EB$ cross modes \cite{Lue:1998mq}.

Using Euler's formula, the relations between polarization power spectra with a constant $\beta$ can be written as \cite{Murai:2022zur}
\begin{equation}
\begin{array}{l}
C_\ell^{E E}=\cos ^2(2 \beta) \tilde{C}_\ell^{E E}+\sin ^2(2 \beta) \tilde{C}_\ell^{B B}  ,
\end{array}
\end{equation}
\begin{equation}
	\begin{array}{l}
		C_\ell^{B B}=\cos ^2(2 \beta) \tilde{C}_\ell^{B B}+\sin ^2(2 \beta) \tilde{C}_\ell^{E E}  ,
	\end{array}
\end{equation}
\begin{equation}
	\begin{array}{l}
		C_\ell^{E B}=\frac{1}{2} \sin (4 \beta)\left(\tilde{C}_\ell^{E E}-\tilde{C}_\ell^{B B}\right),
	\end{array}
 \label{eq:cleb}
\end{equation}
where $\tilde{C}_{\ell}$ are the power spectra in the absence of cosmic birefringence. If we fix $\beta=0$ then $C_\ell^{EE}$ and $C_\ell^{BB}$ reduce to their values in the absence of any photon-scalar coupling and $C_{\ell}^{EB} = 0$. If we assume that the magnitude of $C_\ell^{BB}$ is much smaller than $C_\ell^{E E}$, we can use the approximation $C_\ell^{E B}\approx \tan(2\beta)C_\ell^{E E}$.

Generically, $\beta$ is not constant but rather related to the evolution of the scalar field $\phi$. In this case one must evolve the Boltzmann equations numerically. We will use the publicly available \texttt{CLASS\_{EDE}} program \cite{Hill:2020osr} from \texttt{CLASS}\cite{Lesgourgues:2011re,Blas:2011rf} to solve the Boltzmann equations and obtain the $EE$, $BB$, $EB$ power spectra numerically, for multiple scalar field early dark energy models. Specifically, we focus on two early dark energy models; $\alpha$-attractors and the Rock `n' Roll model, which will be described in the following section.

\section{Early Dark Energy}
\label{sec:3}

Early dark energy is a component of the Universe that introduces a non-negligible energy density during radiation domination. It has been suggested as a possible resolution to the Hubble tension problem by modifying the expansion rate prior to recombination \cite{Caldwell:2003vp, Smith:2019ihp, Berghaus:2019cls, Alexander:2019rsc, Chudaykin:2020acu, Agrawal:2019lmo, Niedermann:2019olb, Freese:2004vs, Ye:2020btb, Akarsu:2019hmw, Lin:2019qug, Yin:2020dwl, Braglia:2020bym}. By construction, early dark energy models contribute energy density during and prior to the epoch of matter-radiation equality, and their effect is to reduce the comoving sound horizon before decoupling. 
 However, these models are difficult to distinguish using current observational data.
 
A recent study \cite{Murai:2022zur} has highlighted that CMB $EB$ cross power can be induced via cosmic birefringence from ultra-light axion-like early dark energy. Hence the $EB$ power spectrum is a new observable that can potentially be used to discriminate different EDE models. In this work, we study EDE models based on pseudo-scalar fields, which differ in their potentials, focusing on their Cosmic Birefringence signals. Before doing so, we will briefly introduce them. 
{The potential of the two cases we used can be considered as pseudo-scalars from the parameters introduced below. }
Practically they share the same features as the axion field -- rolling followed by oscillatory epochs.

\subsection{Early Dark Energy from $\alpha$-attractors}

The $\alpha$-attractors are a class of supergravity-inspired models, first studied within the context of inflation. They have the interesting property that there exists a broad class of scalar field potentials which share a generic prediction for the scalar spectral tilt $n_s$ and tensor-to-scalar ratio $r$, practically independent of the parameters in $V(\phi)$. This property has been dubbed as `attractor' behaviour. This scalar field model has been subsequently adopted as a candidate for early dark energy, simply by reducing the energy scales associated with the scalar field from inflationary to those typically associated with { EDE}.

Defining dynamical variables such that the scalar field kinetic term is canonical, the potential for this class of models can be written as \cite{Linde:2015uga}
\begin{equation}
	V_{\alpha}(\phi)=V_0 \frac{(1+\alpha_2)^{2 n} \tanh \left(\phi / \sqrt{6 \alpha_1} M_{\mathrm{pl}}\right)^{2 p}}{\left[1+\alpha_2 \tanh \left(\phi / \sqrt{6 \alpha_1} M_{\mathrm{pl}}\right)\right]^{2 n}}
	\label{phialpha}
\end{equation}
where $V_0$, $p$, $n$, $\alpha_1$ and $\alpha_2$ are constants. The particular case $\alpha_2=1$ was considered in Ref.\cite{Braglia:2020bym} and based on the numerical results in that work, we select the best-fit case $p=2$ and $n=0$ to compare with other early dark energy models. 
{ Only at $n=0$, the potential has $V(\phi)=V(-\phi)$ and can be calculated as pseudo-scalar.}
We introduce the dimensionless field
$\Theta_{\alpha} \equiv \phi /\left(\sqrt{6 \alpha_1} M_{\mathrm{pl}}\right)$ when presenting the dynamics in the next section. The Taylor expansion of $V(\phi)$ for $\phi \simeq 0$ is given by
$V_{\alpha}(\phi) \simeq \Theta_{\alpha}^4$. 

\subsection{Rock ‘n’ Roll}

The Rock `n’ Roll model introduces an additional scalar field before recombination with potential
\begin{equation}
V_{\rm R\&R}(\phi)=V_0\left(\frac{\phi}{M_{\mathrm{Pl}}}\right)^{2 n}
\label{phirnr}
\end{equation}
where $M_{Pl}$ is the reduced Planck mass, and $V_0$ and $n$ are constants. We follow Ref.\cite{Agrawal:2019lmo} and fix the index $n=2$.

For the Rock `n' Roll model, the scalar field will introduce an additional energy density in a narrow redshift window around recombination, injecting energy and reducing the acoustic horizon. For this reason, the Hubble constant $H_{0}$ inferred from the CMB will be larger. 
The Taylor expansion around $\phi=0$ is trivial -- $V_{\rm R\&R}(\phi)\propto (\frac{\phi}{M_{Pl}})^4$, similar to the $\alpha$-attractor model. The important discriminator for these models is the behaviour of $\phi$ prior to its final state, oscillating in the vicinity of the minima at $\phi \simeq 0$.

Initially, we focus on the evolution of the scalar field and energy density with redshift $z$. The background, homogeneous energy density of the field is simply given by the standard form 

\begin{equation}
	\label{rho}
	\rho_{\phi}=\frac{1}{2} \dot{\phi}^2+V(\phi),
\end{equation}

\noindent since the scalar field is always normalized to ensure a canonical kinetic term. 

\begin{figure}
	\centering
 	\includegraphics[width=0.6\textwidth]{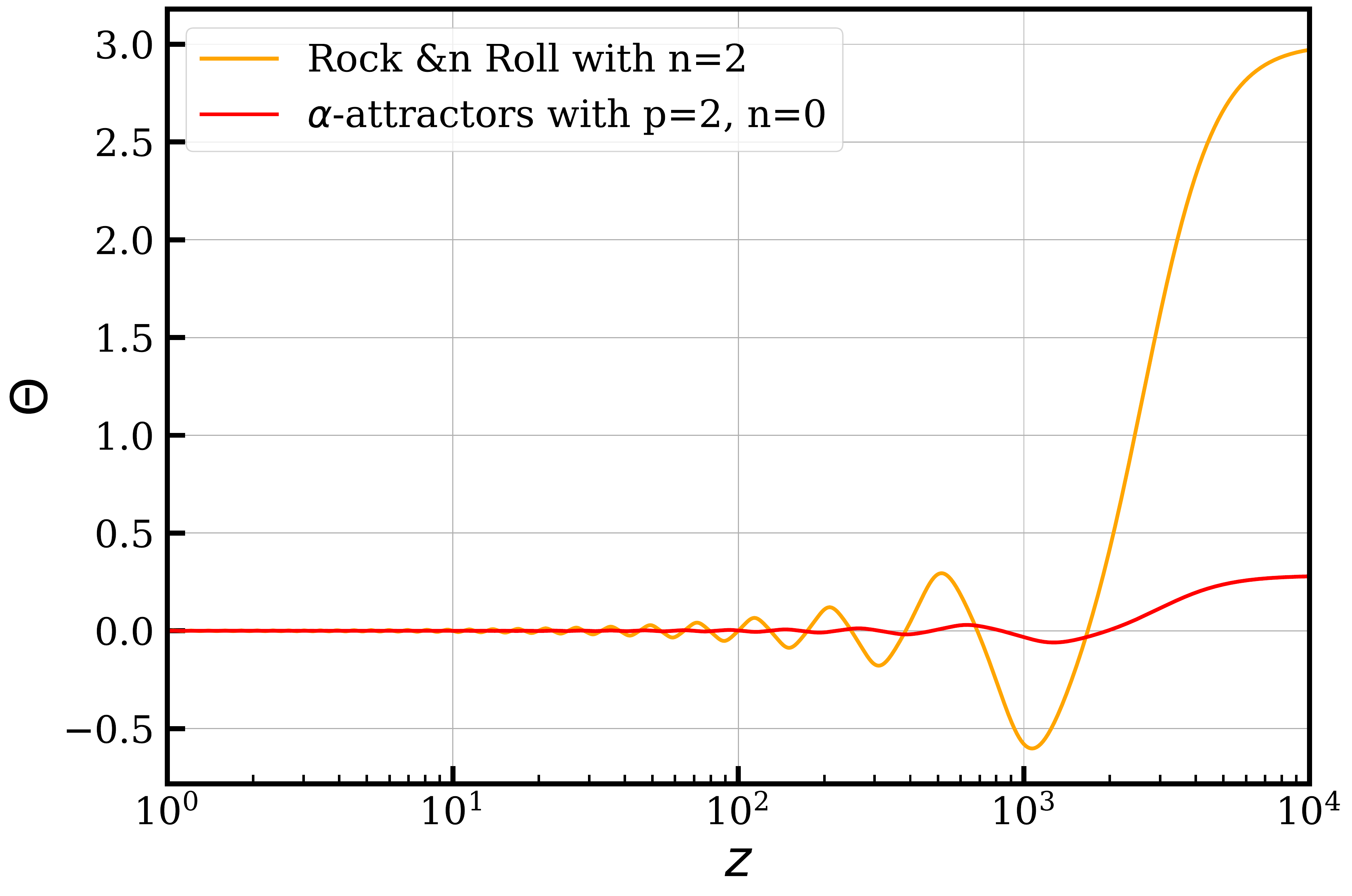}
	\caption{\label{fig:theta}The evolution of scalar field with redshift.}
\end{figure}

In Table \ref{tab:2}, we present the parameter values used in our numerical analysis, which are the best-fit values 
inferred by fitting the models to cosmological data in previous studies using CMB temperature \cite{Poulin:2018cxd, Planck:2019nip, Planck:2018nkj}, Baryon Acoustic Oscillation (BAO) \cite{BOSS:2016wmc, BOSS:2016apd, Vargas-Magana:2016imr, BOSS:2016hvq}, and type Ia Supernova \cite{Pan-STARRS1:2017jku, Riess:2019cxk} data. 
$f_{\rm ede}$ is the maximum energy density fraction of early dark energy at critical redshift $z_c$, which is the time at which the scalar field starts oscillating ($z_c=a_c^{-1}-1$). 
By solving the background cosmological equations for the scalar fields described by potentials (\ref{phialpha}, \ref{phirnr}), using parameter values taken from Table \ref{tab:2}, we can plot the evolution of the fields, which have different initial epochs and subsequent oscillations. The dynamical evolution of the fields, as a function of redshift, are presented in Figure \ref{fig:theta}. 
In this figure, the red/orange lines represent the field oscillation for the $\alpha$-attractor, and Rock `n' Roll model respectively. During the oscillatory phase of the dynamics, the Rock `n' Roll model has a higher frequency compared to the $\alpha$-attractor.

\begin{figure}
	\centering
	\includegraphics[width=0.6\textwidth]{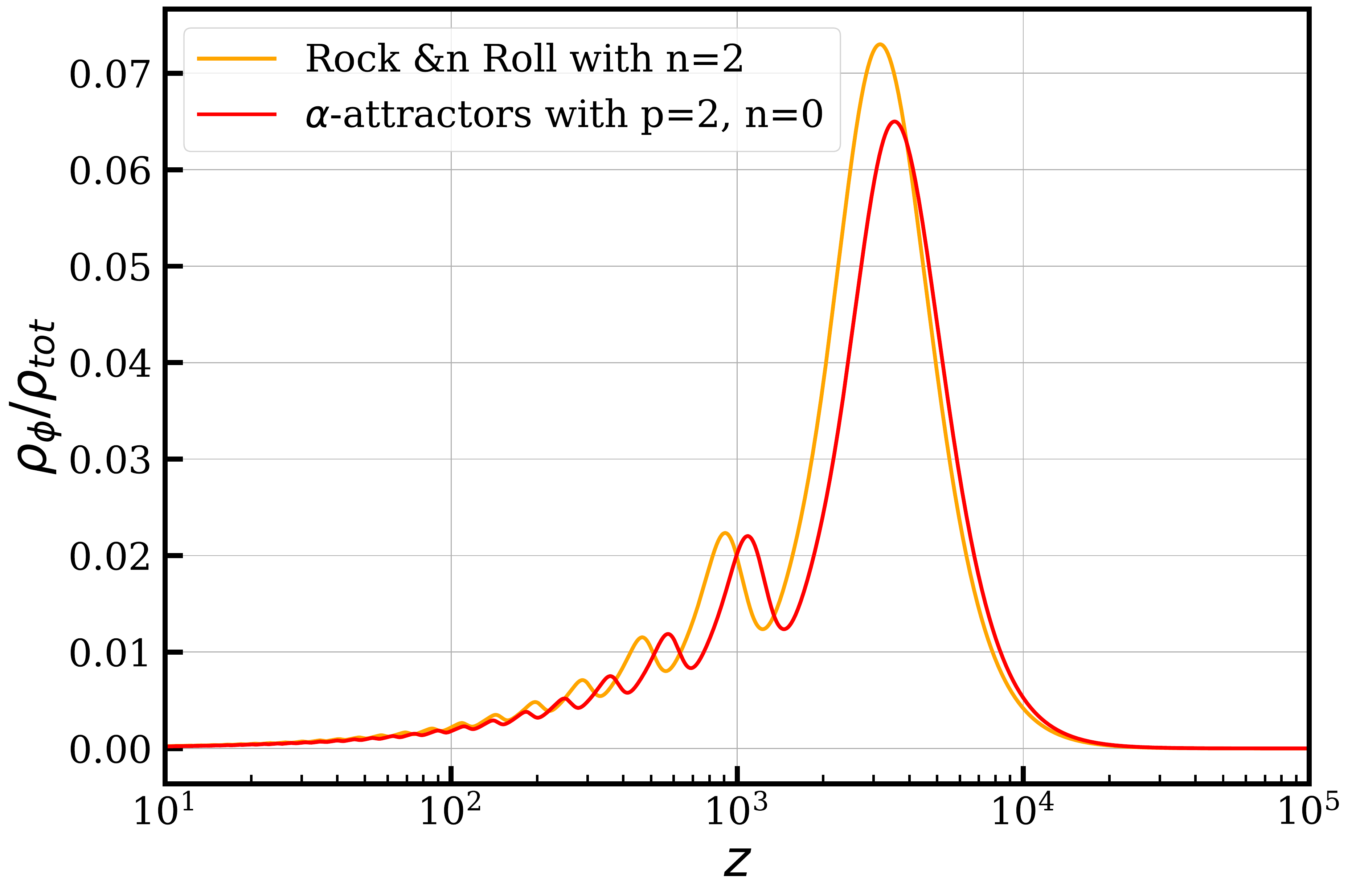}
	\caption{\label{fig:fede}The evolution of $\Omega_{\phi} = 8\pi G \rho_{\phi}/(3H^{2})$ with redshift. 
 }
\end{figure}

The energy density in Equation \ref{rho} for the different models is shown in Figure \ref{fig:fede}. 
The Rock `n' Roll model {has a higher peak value of $\rho_{\phi}$  than $\alpha$-attractor's }prior to the decoupling epoch.
 The redshift of the peak of $\rho_{\phi}$ defines the redshift $z_c$, which is related to the scale factor $\log_{10}(a_c)$ given in Table \ref{tab:2}. Both models present similar energy densities. 

The two models in Figure \ref{fig:fede} possess very similar values of $\rho_{\phi}$ at the decoupling epoch ($z=1100$). After this time, both decay at similar rates. The length of the energy density tail is another potential discriminator between models.
However, both models considered here possess negligible energy density by $z = 10$. 

From Table \ref{tab:2}, we observe that the EDE models can ameliorate the $H_0$ tension, both via a shift in the best-fit value of $H_0$, as well as an increase in the size of the error bars. Both models increase the $H_0$ best fit to a value larger than $70 \, {\rm km \, s}^{-1} \, {\rm Mpc}^{-1}$, which decreases the tension with the low redshift data to $\sim 2\sigma$.

\begin{table}
	\begin{center}
	\caption{{\color{black}Best-fit of free parameters in $\Lambda$CDM, $\alpha$-attractor and Rock `n' Roll models. These results with 68$\%$ C.L. are based on the observation of CMB, BAO, and Type Ia Supernova. 
 }}
	\begin{tabular}{|l|c|c|c|}
		 \hline $\text { Parameter } $&  $\Lambda$CDM          &$ \alpha$-attractor&Rock `n' Roll \\
		\hline $100 \theta_s $&       $ 1.04213 \pm 0.0003 $&$ 1.04154 \pm 0.00037 $&$1.0417\pm0.0004$\\
		$100 \omega_b $&              $ 2.239 \pm 0.014 $   &$ 2.277 \pm 0.024 $&$2.264^{+0.021}_{-0.020}$\\
		$\omega_{\text {cdm }} $&     $ 0.1177 \pm 0.0012 $ &$ 0.1265 \pm 0.0036$& $0.1267^{+0.0044}_{-0.0043}$\\
		$10^9 A_s $&                  $ 2.14 \pm 0.051 $    &$ 2.1477^{+0.03464}_{-0.03409} $& $2.2309^{0.5876}_{0.5072}$\\
		$n_s $&                       $ 0.9687 \pm 0.0044 $ &$ 0.9803 \pm 0.0057 $&$ 0.977^{+0.006}_{-0.007} $\\
		$\tau_{\text {reio }}$&       $ 0.068 \pm 0.013 $   &$ 0.0602^{+0.0070}_{-0.0081} $&$ 0.080^{+0.013}_{-0.012} $\\
		$\log _{10}\left(a_c\right) $&$- $                  & $-3.550_{-0.061}^{+0.074}$&$ -3.4998^{+0.0965}_{-0.0431}$ \\
		$f_{\mathrm{EDE}}\left(a_c\right)$& $- $            & $0.065\pm{0.026} $&$ 0.073^{+0.031}_{-0.028} $\\
  		{$H_0$}&                      $68.33\pm 0.54 $      & $70.28\pm 0.94 $&$ 70.5^{+1.0}_{-1.2}$\\
		\hline
 		\end{tabular}
\label{tab:2}
\end{center}
\end{table}

\section{Cosmic Birefringence from $\alpha$-attractors and Rock `n' Roll Early Dark Energy}
\label{sec:4}

\begin{figure}
\centering
\includegraphics[width=1\textwidth]{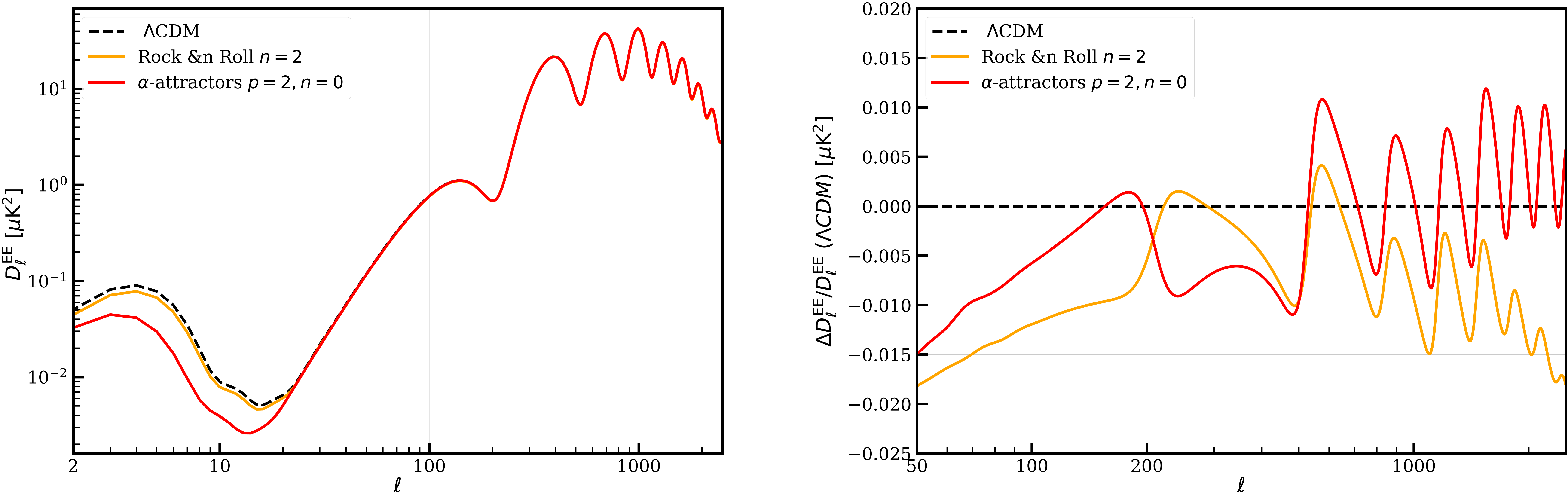}
\caption{\label{fig:EEBB}CMB power spectra of $EE$ mode. The early dark energy models both include an effect due to cosmic birefringence. 
}
\end{figure}

\begin{figure}
\centering
\includegraphics[width=0.6\textwidth]{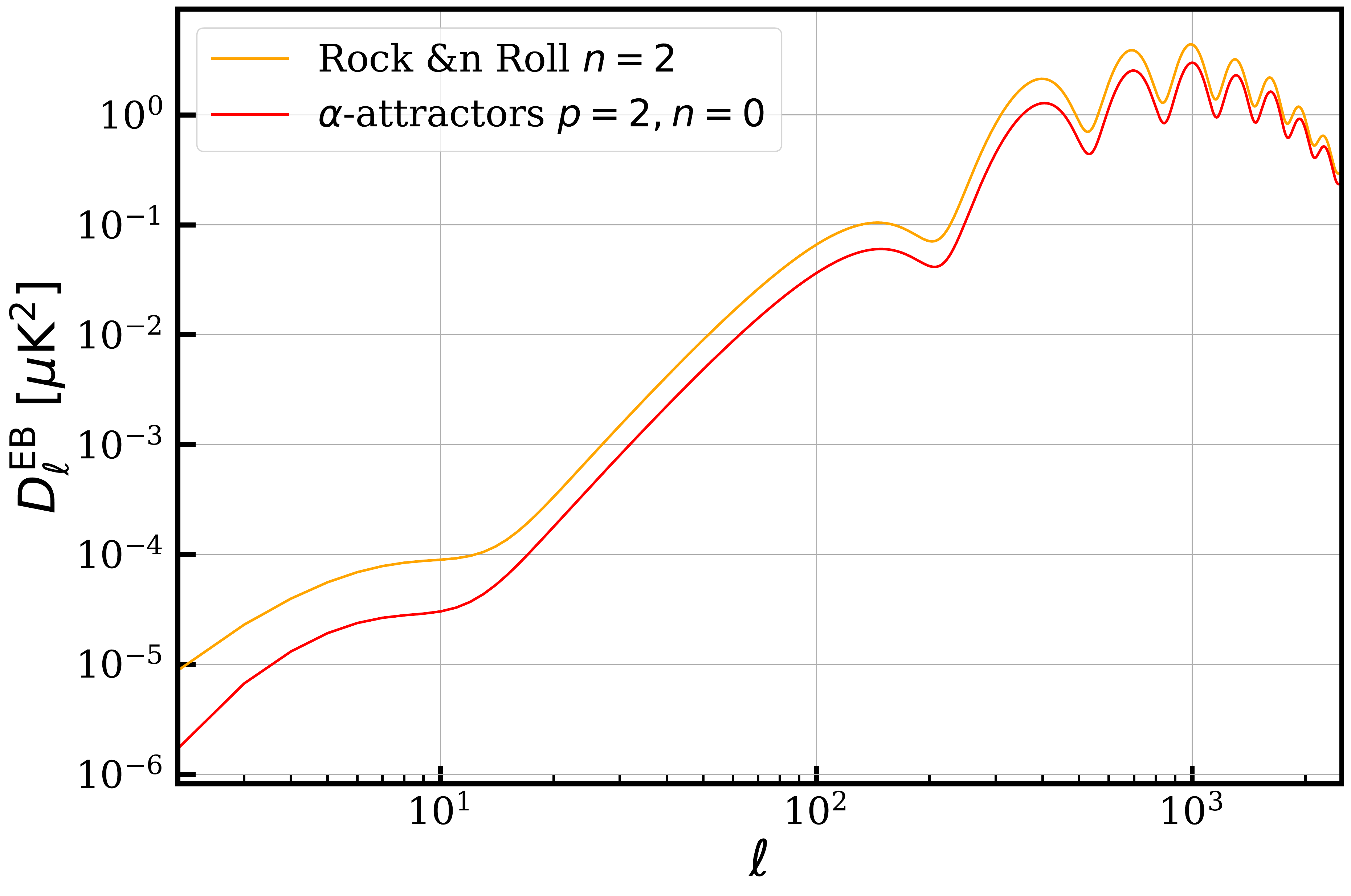}
\caption{\label{fig:EB}$EB$ power spectrum in the four cases with best-fitting parameters given in Table \ref{tab:2}. The value of $gM_{pl}$ we used in this figure is 1. }
\end{figure}

\begin{figure}
	\centering
	\includegraphics[width=0.6\textwidth]{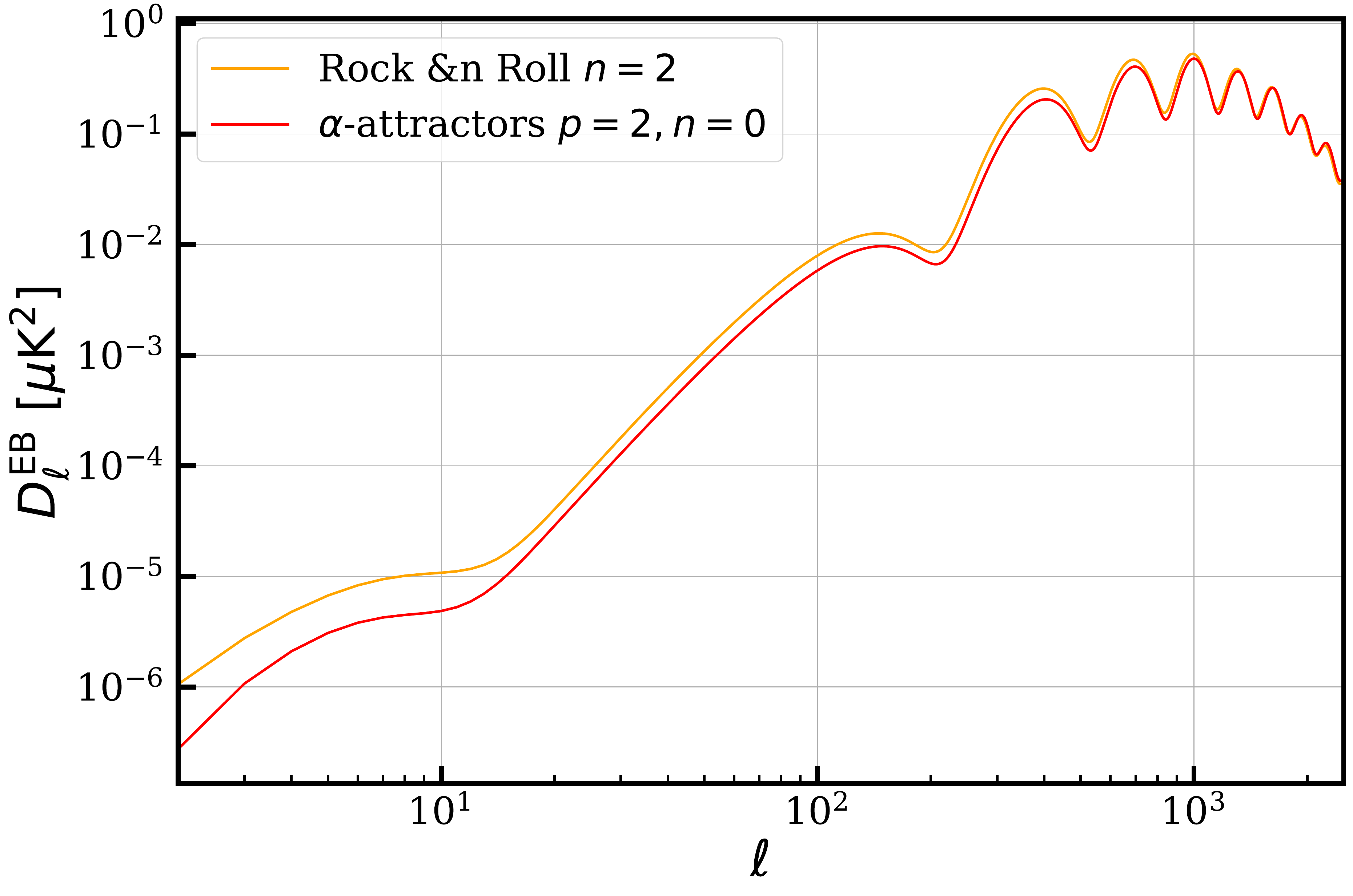}
	\caption{\label{fig:EB-g}With best fitted  $gM_{Pl} = 0.16$ in $\alpha$-attractor and $gM_{Pl} = 0.12$ in Rock `n' Roll, $EB$ power spectrum in the two cases can be modified from the above result.}
\end{figure}

\begin{figure}
\centering
\includegraphics[width=0.49\textwidth]{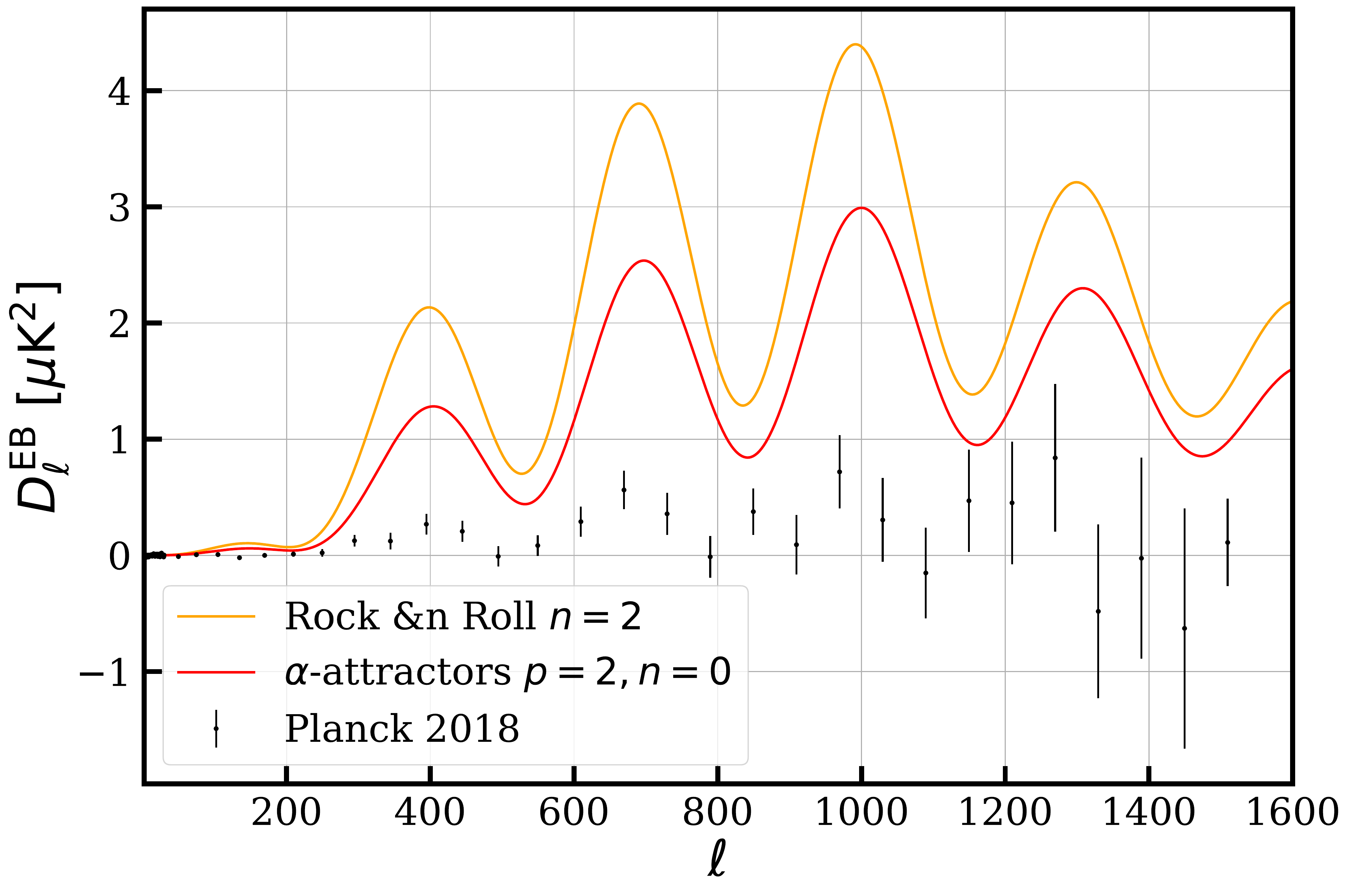}
\includegraphics[width=0.49\textwidth]{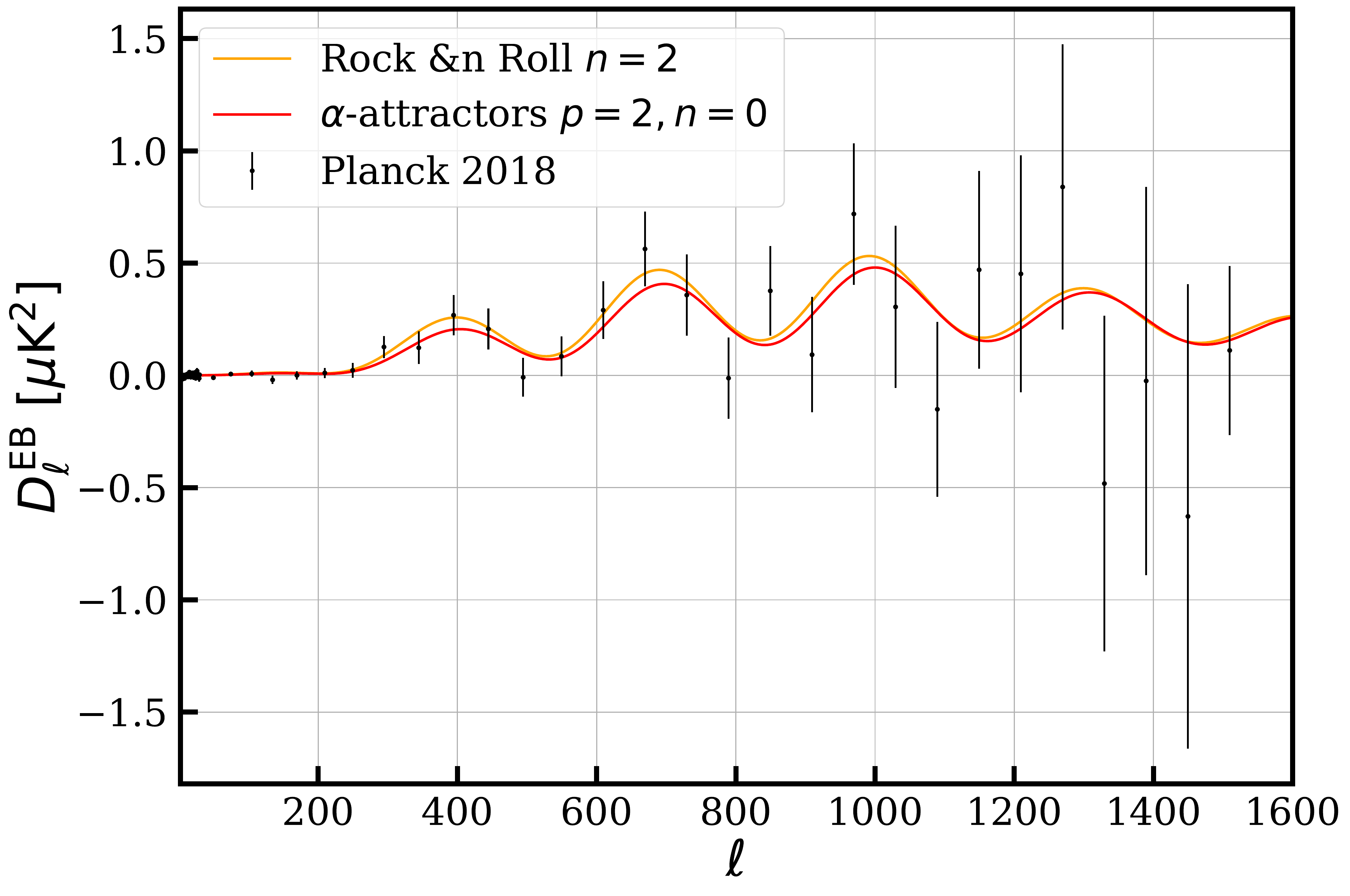}
\caption{\label{fig:EB_planck}$EB$ power spectrum with the data points from Planck 2018. {The left figure plotted with $gM_{Pl}=1$, and the right one shows the $EB$ mode with best-fitted $gM_{Pl}$ from Table. \ref{tab:g}. }}
\end{figure}


\begin{figure}
\centering
\includegraphics[width=0.6\textwidth]{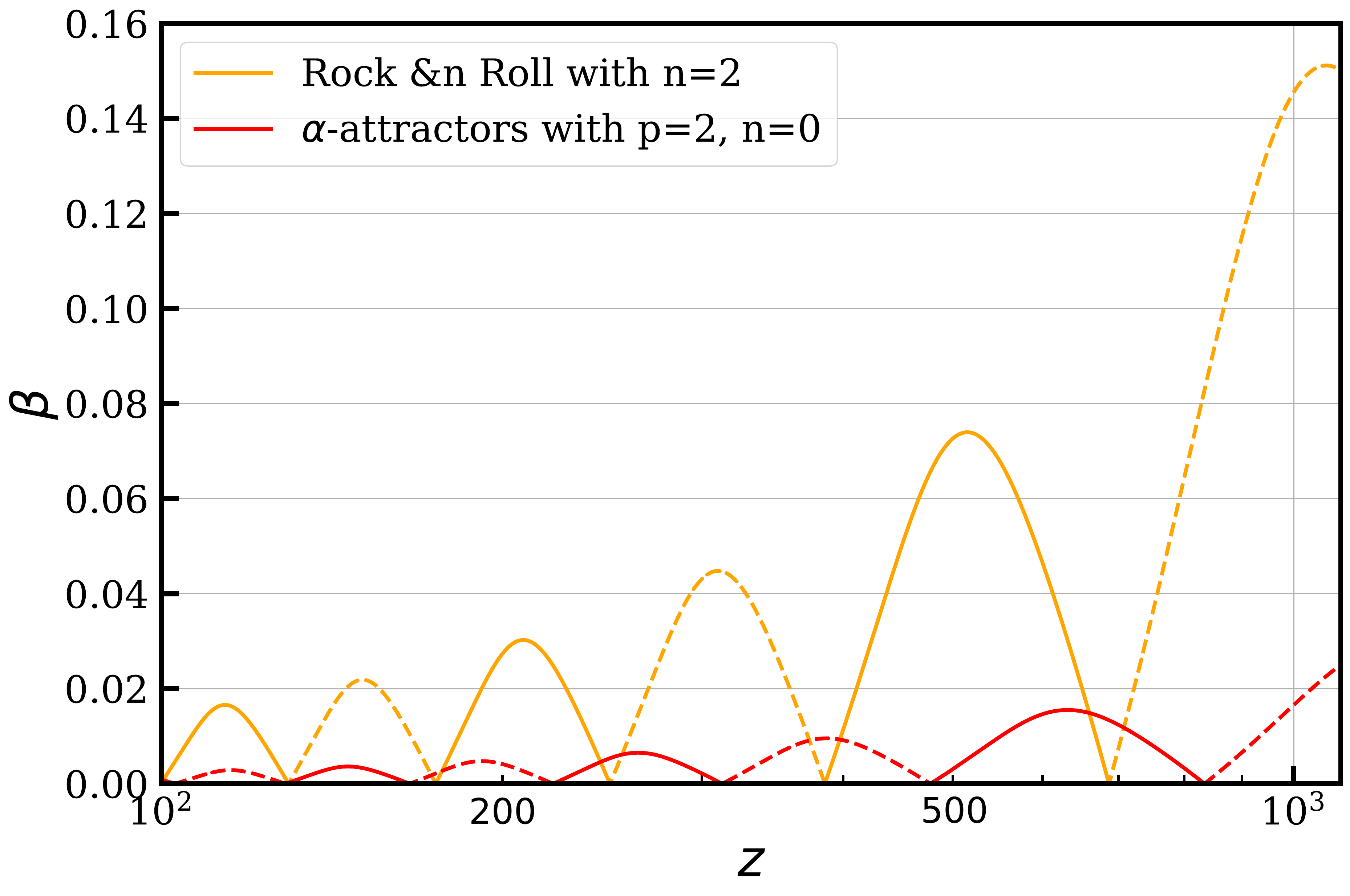}
\caption{\label{fig:EB-best-g}The evolution of $\beta$ with redshift.}
\label{fig:beta}
\end{figure}

In this section, we study the effect of early dark energy -- with explicit Chern-Simons coupling to photons --  on the Boltzmann equation, accounting for the dynamical nature of $\phi$ in Equation \ref{Boltzmann}. We implement the dynamics of these different models into \texttt{CLASS\_{EDE}} \cite{Hill:2020osr}
 and compare the resulting CMB power spectra. We include the gravitational lensing effect on the CMB. More discussion on the lensing effect can be found in Ref.\cite{Murai:2022zur}.

{\subsection{Coupling with $gM_{Pl}=1$}}

We focus on the $EE$ and $EB$ power spectra. Since the initial evolution and subsequent oscillation frequency of $\phi$ is different for each model, $\beta(\phi)$ will change both the amplitude and shape of these power spectra. 
The $EE$ power spectra are presented in the left panel of Figure \ref{fig:EEBB}, and the fractional difference between early dark energy and $\Lambda$CDM power spectra in the right panel. The two cases present different amounts of power in the low-$l$ scales, and almost coincide in the high-$\ell$ period. This is predominantly due to the different values of $a_c$ in each model.
If $a_c$ is closer to the reionization period, then observations of sources at different redshifts, such as photons from CMB and reionization, will give a different rotation angle. Thus, we can extract information about the redshift evolution of $\beta$ by looking at sources in different redshift shells, which will provide improved constraints on this class of models.  
Although the low-$\ell$ $EE$ power differs significantly between the different models, the large statistical uncertainties in the Planck data in this regime make it difficult to distinguish different models.
The $\alpha$-attractor has less $EE$ power for $\ell <26$. On smaller angular scales, the differences between the spectra are smaller, but the error bars are also expected to be small for $\ell \gtrsim 10^{2}$. 

Next we focus on the $EB$ power spectra for these early dark energy models, presented in Figure \ref{fig:EB}. For the Rock `n' Roll model, the $EB$ power spectrum has {a larger amplitude } in both the low-$\ell$ and high-$\ell$ regimes. This is a direct consequence of $f_{\rm ede}$ taking { a larger} value in the Rock `n' Roll model. 
In contrast, { both the $EB$ power spectra and the low-$\ell$ component of the $EE$ power spectra  show such positive correlation relation} between $f_{\rm ede}$ and power. The magnitude of $D_\ell^{EE}$ for $\ell < 26$ is suppressed for the $\alpha$-attractor. 

At high-$\ell$ modes, there is also a subtle difference in shape of the $EB$ power spectra. The Rock `n' Roll power spectrum is higher than the $\alpha$-attractor counterpart at $l=1000$ but decays faster at larger $\ell$. The peak value of Rock `n' Roll also shifts slightly to the smaller scales. 

{ To test the viability of both the models, we use data from the Planck 2018 release. First, we compute the pseudo-cl $EB$ spectra from the maps using the Planck common polarization mask with 2$^\circ$ apodization. Then we estimate the $EB$ power spectra by correcting for the effects of masking, beam and pixel windows. The $EB$ spectra that we obtain are comparable to those in \cite{Eskilt:2022cff}. We check the stability of our results with changes to the sky fraction by using two different masks, the SMICA confidence mask and the Planck common polarization mask. The SMICA confidence mask has a sky fraction of 93\%, while the common mask has a sky fraction of 78\%. We find that both the masks yield similar $EB$ power spectra, and so we use the Planck common polarization mask for our analysis.}



{In the left panel of Figure.\ref{fig:EB_planck},
we add the result from observation into the $EB$ power spectra and want to use it  to distinguish early dark energy models from $l=200$ to $l=1500$, although the large errorbars make it hard to distinguish the models at low $\ell$s. But this is the result from fixing the coupling term $g$ as $g/M^{-1}_{Pl}=1$. 
In the next, we will consider the value of $gM_{Pl}$ from the best fit with Planck data.}

\begin{table}
	\begin{center}
	\caption{{\color{black}Best-fit of free parameters of $gM_{Pl}$ and the corresponding value of $\beta$ in $\Lambda$CDM,  $\alpha$-attractor, Rock `n' Roll, respectively. Since in our theory, $\beta$ is not constant, we only give the value of $\beta$ calculated from Equation\ref{beta} in CMB time.
 }}
	\begin{tabular}{|l|c|c|c|}
		 \hline $\text { Parameter } $& $\Lambda$CDM&$ \alpha$-attractor&Rock `n' Roll \\
		\hline $gM_{Pl} $&$ 0$&$ 0.16 $&$0.12$\\
  \hline
  \hline
		$\beta $ at CMB &$ 0 $&$ 0.02^{\circ} $&$0.15^{\circ}$\\
		\hline
 		\end{tabular}
\label{tab:g}
\end{center}
\end{table}

\subsection{Coupling term $g$ from best-fit result}

The amplitude of the $EB$ power spectrum is dictated by a single free parameter, $g$.
In this section, we fit the parameter $g$ for both EDE models. The best-fit values of $g M_{Pl}$ for both are presented in Table. \ref{tab:g}. We {find the results of $gM_{Pl}$ are} $g = 0.16$($\alpha$-attractors) and $g = 0.12$ (Rock n Roll), respectively.
In {the right one of Figure \ref{fig:EB_planck}, we present the best-fit results of the models considered here and compare them with the $EB$ spectra measured by Planck. To fit $g$, we used parameters except $g$  from Table \ref{tab:2} obtained by fitting the CMB \cite{Poulin:2018cxd, Planck:2019nip, Planck:2018nkj}, BAO\cite{BOSS:2016wmc, BOSS:2016apd, Vargas-Magana:2016imr, BOSS:2016hvq} and Supernova \cite{Pan-STARRS1:2017jku, Riess:2019cxk} data.} Then for each model, we vary $g$ find the best fit value of $g$ that minimizes the $\chi^2$; 

\begin{equation} \chi^2(g) = \sum_\ell {\left(\mathcal{D}_\ell^O - \mathcal{D}_\ell^M(g)\right)^2 \over \sigma_\ell^2 } ,
\end{equation} 

\noindent where $\mathcal{D}_\ell^O$ and $\mathcal{D}_\ell^M(g)$ are the measured $EB$ spectra and the model predictions, respectively, and $\sigma_\ell$ are the Planck $EB$ 68\% error bars. We find that both models can fit the data, and the differences between the best fits from the two models cannot be distinguished by the present data at any statistical significance.


We present the $EB$ power spectra for both models with best-fit $g$ in Figure \ref{fig:EB-g}. This result shows the $EB$ power spectrum cannot distinguish the two models when $g$ is allowed to vary. {For higher $l >1500$,} the EB mode will show very subtle differences, but the shape is practically indistinguishable. 

We use the best-fit value of $g$ with Equation \ref{beta} to infer value of $\beta$. The results {also be shown in} Table. \ref{tab:g} and Figure \ref{fig:beta}. With the best fit $g$, we find the rotation angle $\beta$ is different for the two models: the $\alpha$-attractor model with $g=0.16$ can fit the data of $EB$ mode well and possess a relatively small { rotation angle $\beta=0.02^{\circ}$,  and the Rock `n' Roll model with $gM_{Pl}=0.12$ can get $\beta$ equal to $0.15^{\circ}$} We conclude that the rotating angle $\beta$ is strongly model dependent.   

These features could help the observation of LiteBIRD distinguish different early energy models.
It may also be the case that not all oscillatory potentials will lead to the oscillation of $D_\ell^{EB}$, since this observable relies on the parameters in the scalar field potential. 

\section{Summary}
\label{sec:5}

In this article, we studied the cosmic birefringence effect from two models; Rock `n' Roll, and $\alpha$-attractor scalar fields, treating them as pseudo-scalars and including an explicit Chern-Simons coupling between the field and photons. We generate the resulting $EB$ power spectrum for the $\alpha$-attractor and Rock `n' Roll models for the first time. We compared the field oscillations and background energy density evolution. Both models admit similar energy density evolution.

We also discussed the cosmic birefringence effect wrought by different EDE potentials.
The ${EE}$ power spectra show some discriminatory power in the low-$\ell$ regime, but the current generation of polarization data can't take advantage of this difference. Cosmic variance will ultimately limit the utility of this signal. Oscillations in the $EE$ power spectrum at high $\ell$ could act as a model discriminant in the next generation of polarization experiments. The signal in the $EE$ power spectrum is present even in the absence of any explicit photon-pseudoscalar coupling. 

The $EB$ cross power shows some promise to distinguish different scalar field models. 
At high-$\ell$, the magnitude of $D_\ell^{EB}$ is closely related to the value of $f_{\rm ede}$, required to resolve the Hubble tension. This implies that the higher the proportion of dark energy density in the early Universe, the greater signal we expect to find in the $EB$ cross-power spectrum. 
At low-$\ell$, both the shape and amplitude of $D_\ell^{EB}$ depend on the model specifics of the model, allowing for stronger discriminatory power. However, using the best fit EDE parameter values from other data sets, in conjunction with fitting the amplitude of the $EB$ power spectrum to the models in this work reveals that both the $\alpha$-attractors and the Rock n' Roll models fit the data equally well. The value of $g$ is model dependent. Moreover, the rotation angle $\beta$ calculated for each model with the best-fit $g$ is also highly model dependent. 

We conclude that the CMB $EB$ power produced by cosmic birefringence could potentially be an important smoking gun for different early dark energy models, beyond the $EE$ spectra, and this will be verified with the next generation of CMB observations in the near future.

\section*{Acknowledgements}

We thank Eiichiro Komatsu, Kai Murai, and  Thomas Tram for their kind help and discussion in numerical calculation. We also thank Stephen Appleby for discussions and his comments on the manuscript.
LY is supported by an appointment to the YST Program at the APCTP through the Science and Technology Promotion Fund and Lottery Fund of the Korean Government. JK is supported by an appointment to the JRG Program at the APCTP through the Science and Technology Promotion Fund and Lottery Fund of the Korean Government, and also supported by the Korean Local Governments in Gyeongsangbuk-do Province and Pohang City.
This research was also supported by the National Research Foundation of Korea  funded through the CQUeST (NRF-2020R1A6A1A03047877), and the Ministry of Science and ICT to BHL (NRF-2020R1F1A1075472).


\begin{thebibliography}{100}
	
	
\bibitem{Komatsu:2014ioa}
E.~Komatsu \textit{et al.} [WMAP Science Team],
PTEP \textbf{2014}, 06B102 (2014)
doi:10.1093/ptep/ptu083
[arXiv:1404.5415 [astro-ph.CO]].

\bibitem{Planck:2018vyg}
N.~Aghanim \textit{et al.} [Planck],
Astron. Astrophys. \textbf{641}, A6 (2020)
[erratum: Astron. Astrophys. \textbf{652}, C4 (2021)]
doi:10.1051/0004-6361/201833910
[arXiv:1807.06209 [astro-ph.CO]].

\bibitem{SPT-3G:2021eoc}
D.~Dutcher \textit{et al.} [SPT-3G],
Phys. Rev. D \textbf{104}, no.2, 022003 (2021)
doi:10.1103/PhysRevD.104.022003
[arXiv:2101.01684 [astro-ph.CO]].

\bibitem{BICEP:2021xfz}
P.~A.~R.~Ade \textit{et al.} [BICEP and Keck],
Phys. Rev. Lett. \textbf{127}, no.15, 151301 (2021)
doi:10.1103/PhysRevLett.127.151301
[arXiv:2110.00483 [astro-ph.CO]].

\bibitem{SPIDER:2021ncy}
P.~A.~R.~Ade \textit{et al.} [SPIDER],
Astrophys. J. \textbf{927}, no.2, 174 (2022)
doi:10.3847/1538-4357/ac20df
[arXiv:2103.13334 [astro-ph.CO]].

\bibitem{Abdalla:2022yfr}
E.~Abdalla, G.~Franco Abell\'an, A.~Aboubrahim, A.~Agnello, O.~Akarsu, Y.~Akrami, G.~Alestas, D.~Aloni, L.~Amendola and L.~A.~Anchordoqui, \textit{et al.}
JHEAp \textbf{34}, 49-211 (2022)
doi:10.1016/j.jheap.2022.04.002
[arXiv:2203.06142 [astro-ph.CO]].

\bibitem{Komatsu:2022nvu}
E.~Komatsu,
Nature Rev. Phys. \textbf{4}, no.7, 452-469 (2022)
doi:10.1038/s42254-022-00452-4
[arXiv:2202.13919 [astro-ph.CO]].

\bibitem{Bernal:2016gxb}
J.~L.~Bernal, L.~Verde and A.~G.~Riess,
JCAP \textbf{10}, 019 (2016)
doi:10.1088/1475-7516/2016/10/019
[arXiv:1607.05617 [astro-ph.CO]].

\bibitem{Carroll:1989vb}
S.~M.~Carroll, G.~B.~Field and R.~Jackiw,
Phys. Rev. D \textbf{41}, 1231 (1990)
doi:10.1103/PhysRevD.41.1231

\bibitem{Carroll:1991zs}
S.~M.~Carroll and G.~B.~Field,
Phys. Rev. D \textbf{43}, 3789 (1991)
doi:10.1103/PhysRevD.43.3789

\bibitem{Harari:1992ea}
D.~Harari and P.~Sikivie,
Phys. Lett. B \textbf{289}, 67-72 (1992)
doi:10.1016/0370-2693(92)91363-E

\bibitem{Minami:2020odp}
Y.~Minami and E.~Komatsu,
Phys. Rev. Lett. \textbf{125}, no.22, 221301 (2020)
doi:10.1103/PhysRevLett.125.221301
[arXiv:2011.11254 [astro-ph.CO]].

\bibitem{Eskilt:2022cff}
J.~R.~Eskilt and E.~Komatsu,
Phys. Rev. D \textbf{106}, no.6, 063503 (2022)
doi:10.1103/PhysRevD.106.063503
[arXiv:2205.13962 [astro-ph.CO]].

\bibitem{Lue:1998mq}
A.~Lue, L.~M.~Wang and M.~Kamionkowski,
Phys. Rev. Lett. \textbf{83}, 1506-1509 (1999)
doi:10.1103/PhysRevLett.83.1506
[arXiv:astro-ph/9812088 [astro-ph]].

\bibitem{Nakatsuka:2022epj}
H.~Nakatsuka, T.~Namikawa and E.~Komatsu,
Phys. Rev. D \textbf{105}, no.12, 123509 (2022)
doi:10.1103/PhysRevD.105.123509
[arXiv:2203.08560 [astro-ph.CO]].

\bibitem{Murai:2022zur}
K.~Murai, F.~Naokawa, T.~Namikawa and E.~Komatsu,
Phys. Rev. D \textbf{107}, no.4, L041302 (2023)
doi:10.1103/PhysRevD.107.L041302
[arXiv:2209.07804 [astro-ph.CO]].

\bibitem{SimonsObservatory:2018koc}
P.~Ade \textit{et al.} [Simons Observatory],
JCAP \textbf{02}, 056 (2019)
doi:10.1088/1475-7516/2019/02/056
[arXiv:1808.07445 [astro-ph.CO]].

\bibitem{Abazajian:2019eic}
K.~Abazajian, G.~Addison, P.~Adshead, Z.~Ahmed, S.~W.~Allen, D.~Alonso, M.~Alvarez, A.~Anderson, K.~S.~Arnold and C.~Baccigalupi, \textit{et al.}
[arXiv:1907.04473 [astro-ph.IM]].

\bibitem{LiteBIRD:2022cnt}
E.~Allys \textit{et al.} [LiteBIRD],
doi:10.1093/ptep/ptac150
[arXiv:2202.02773 [astro-ph.IM]].

\bibitem{Riess:2021jrx}
A.~G.~Riess, W.~Yuan, L.~M.~Macri, D.~Scolnic, D.~Brout, S.~Casertano, D.~O.~Jones, Y.~Murakami, L.~Breuval and T.~G.~Brink, \textit{et al.}
Astrophys. J. Lett. \textbf{934}, no.1, L7 (2022)
doi:10.3847/2041-8213/ac5c5b
[arXiv:2112.04510 [astro-ph.CO]].

\bibitem{Caldwell:2003vp}
R.~R.~Caldwell, M.~Doran, C.~M.~Mueller, G.~Schafer and C.~Wetterich,
Astrophys. J. Lett. \textbf{591}, L75-L78 (2003)
doi:10.1086/376975
[arXiv:astro-ph/0302505 [astro-ph]].

\bibitem{Smith:2019ihp}
T.~L.~Smith, V.~Poulin and M.~A.~Amin,
Phys. Rev. D \textbf{101}, no.6, 063523 (2020)
doi:10.1103/PhysRevD.101.063523
[arXiv:1908.06995 [astro-ph.CO]].

\bibitem{Berghaus:2019cls}
K.~V.~Berghaus and T.~Karwal,
Phys. Rev. D \textbf{101}, no.8, 083537 (2020)
doi:10.1103/PhysRevD.101.083537
[arXiv:1911.06281 [astro-ph.CO]].

\bibitem{Alexander:2019rsc}
S.~Alexander and E.~McDonough,
Phys. Lett. B \textbf{797}, 134830 (2019)
doi:10.1016/j.physletb.2019.134830
[arXiv:1904.08912 [astro-ph.CO]].

\bibitem{Chudaykin:2020acu}
A.~Chudaykin, D.~Gorbunov and N.~Nedelko,
JCAP \textbf{08}, 013 (2020)
doi:10.1088/1475-7516/2020/08/013
[arXiv:2004.13046 [astro-ph.CO]].

\bibitem{Agrawal:2019lmo}
P.~Agrawal, F.~Y.~Cyr-Racine, D.~Pinner and L.~Randall,
[arXiv:1904.01016 [astro-ph.CO]].

\bibitem{Niedermann:2019olb}
F.~Niedermann and M.~S.~Sloth,
Phys. Rev. D \textbf{103}, no.4, L041303 (2021)
doi:10.1103/PhysRevD.103.L041303
[arXiv:1910.10739 [astro-ph.CO]].

\bibitem{Freese:2004vs}
K.~Freese and D.~Spolyar,
JCAP \textbf{07}, 007 (2005)
doi:10.1088/1475-7516/2005/07/007
[arXiv:hep-ph/0412145 [hep-ph]].

\bibitem{Ye:2020btb}
G.~Ye and Y.~S.~Piao,
Phys. Rev. D \textbf{101}, no.8, 083507 (2020)
doi:10.1103/PhysRevD.101.083507
[arXiv:2001.02451 [astro-ph.CO]].

\bibitem{Akarsu:2019hmw}
\"O.~Akarsu, J.~D.~Barrow, L.~A.~Escamilla and J.~A.~Vazquez,
Phys. Rev. D \textbf{101}, no.6, 063528 (2020)
doi:10.1103/PhysRevD.101.063528
[arXiv:1912.08751 [astro-ph.CO]].

\bibitem{Lin:2019qug}
M.~X.~Lin, G.~Benevento, W.~Hu and M.~Raveri,
Phys. Rev. D \textbf{100}, no.6, 063542 (2019)
doi:10.1103/PhysRevD.100.063542
[arXiv:1905.12618 [astro-ph.CO]].

\bibitem{Yin:2020dwl}
L.~Yin,
Eur. Phys. J. C \textbf{82}, no.1, 78 (2022)
doi:10.1140/epjc/s10052-022-10020-w
[arXiv:2012.13917 [astro-ph.CO]].

\bibitem{Braglia:2020bym}
M.~Braglia, W.~T.~Emond, F.~Finelli, A.~E.~Gumrukcuoglu and K.~Koyama,
Phys. Rev. D \textbf{102}, no.8, 083513 (2020)
doi:10.1103/PhysRevD.102.083513
[arXiv:2005.14053 [astro-ph.CO]].

\bibitem{Hill:2020osr}
J.~C.~Hill, E.~McDonough, M.~W.~Toomey and S.~Alexander,
Phys. Rev. D \textbf{102}, no.4, 043507 (2020)
doi:10.1103/PhysRevD.102.043507
[arXiv:2003.07355 [astro-ph.CO]].

\bibitem{Lesgourgues:2011re}
J.~Lesgourgues,
[arXiv:1104.2932 [astro-ph.IM]].

\bibitem{Blas:2011rf}
D.~Blas, J.~Lesgourgues and T.~Tram,
JCAP \textbf{07}, 034 (2011)
doi:10.1088/1475-7516/2011/07/034
[arXiv:1104.2933 [astro-ph.CO]].

\bibitem{Witten:1984dg}
E.~Witten,
Phys. Lett. B \textbf{149}, 351-356 (1984)
doi:10.1016/0370-2693(84)90422-2

\bibitem{Svrcek:2006yi}
P.~Svrcek and E.~Witten,
JHEP \textbf{06}, 051 (2006)
doi:10.1088/1126-6708/2006/06/051
[arXiv:hep-th/0605206 [hep-th]].

\bibitem{Douglas:2006es}
M.~R.~Douglas and S.~Kachru,
Rev. Mod. Phys. \textbf{79}, 733-796 (2007)
doi:10.1103/RevModPhys.79.733
[arXiv:hep-th/0610102 [hep-th]].

\bibitem{Arvanitaki:2009fg}
A.~Arvanitaki, S.~Dimopoulos, S.~Dubovsky, N.~Kaloper and J.~March-Russell,
Phys. Rev. D \textbf{81}, 123530 (2010)
doi:10.1103/PhysRevD.81.123530
[arXiv:0905.4720 [hep-th]].

\bibitem{Marsh:2015xka}
D.~J.~E.~Marsh,
Phys. Rept. \textbf{643}, 1-79 (2016)
doi:10.1016/j.physrep.2016.06.005
[arXiv:1510.07633 [astro-ph.CO]].

\bibitem{Choi:2006qj}
K.~S.~Choi, I.~W.~Kim and J.~E.~Kim,
JHEP \textbf{03}, 116 (2007)
doi:10.1088/1126-6708/2007/03/116
[arXiv:hep-ph/0612107 [hep-ph]].

\bibitem{Choi:2009jt}
K.~S.~Choi, H.~P.~Nilles, S.~Ramos-Sanchez and P.~K.~S.~Vaudrevange,
Phys. Lett. B \textbf{675}, 381-386 (2009)
doi:10.1016/j.physletb.2009.04.028
[arXiv:0902.3070 [hep-th]].

\bibitem{Poulin:2018dzj}
V.~Poulin, T.~L.~Smith, D.~Grin, T.~Karwal and M.~Kamionkowski,
Phys. Rev. D \textbf{98}, no.8, 083525 (2018)
doi:10.1103/PhysRevD.98.083525
[arXiv:1806.10608 [astro-ph.CO]].

\bibitem{Linde:2015uga}
A.~Linde,
JCAP \textbf{05}, 003 (2015)
doi:10.1088/1475-7516/2015/05/003
[arXiv:1504.00663 [hep-th]].

\bibitem{Poulin:2018cxd}
V.~Poulin, T.~L.~Smith, T.~Karwal and M.~Kamionkowski,
Phys. Rev. Lett. \textbf{122}, no.22, 221301 (2019)
doi:10.1103/PhysRevLett.122.221301
[arXiv:1811.04083 [astro-ph.CO]].

\bibitem{Planck:2019nip}
N.~Aghanim \textit{et al.} [Planck],
Astron. Astrophys. \textbf{641}, A5 (2020)
doi:10.1051/0004-6361/201936386
[arXiv:1907.12875 [astro-ph.CO]].

\bibitem{Planck:2018nkj}
N.~Aghanim \textit{et al.} [Planck],
Astron. Astrophys. \textbf{641}, A1 (2020)
doi:10.1051/0004-6361/201833880
[arXiv:1807.06205 [astro-ph.CO]].

\bibitem{BOSS:2016wmc}
S.~Alam \textit{et al.} [BOSS],
Mon. Not. Roy. Astron. Soc. \textbf{470}, no.3, 2617-2652 (2017)
doi:10.1093/mnras/stx721
[arXiv:1607.03155 [astro-ph.CO]].

\bibitem{BOSS:2016apd}
A.~J.~Ross \textit{et al.} [BOSS],
Mon. Not. Roy. Astron. Soc. \textbf{464}, no.1, 1168-1191 (2017)
doi:10.1093/mnras/stw2372
[arXiv:1607.03145 [astro-ph.CO]].

\bibitem{Vargas-Magana:2016imr}
M.~Vargas-Maga\~na, S.~Ho, A.~J.~Cuesta, R.~O'Connell, A.~J.~Ross, D.~J.~Eisenstein, W.~J.~Percival, J.~N.~Grieb, A.~G.~S\'anchez and J.~L.~Tinker, \textit{et al.}
Mon. Not. Roy. Astron. Soc. \textbf{477}, no.1, 1153-1188 (2018)
doi:10.1093/mnras/sty571
[arXiv:1610.03506 [astro-ph.CO]].

\bibitem{BOSS:2016hvq}
F.~Beutler \textit{et al.} [BOSS],
Mon. Not. Roy. Astron. Soc. \textbf{464}, no.3, 3409-3430 (2017)
doi:10.1093/mnras/stw2373
[arXiv:1607.03149 [astro-ph.CO]].

\bibitem{Pan-STARRS1:2017jku}
D.~M.~Scolnic \textit{et al.} [Pan-STARRS1],
Astrophys. J. \textbf{859}, no.2, 101 (2018)
doi:10.3847/1538-4357/aab9bb
[arXiv:1710.00845 [astro-ph.CO]].

\bibitem{Riess:2019cxk}
A.~G.~Riess, S.~Casertano, W.~Yuan, L.~M.~Macri and D.~Scolnic,
Astrophys. J. \textbf{876}, no.1, 85 (2019)
doi:10.3847/1538-4357/ab1422
[arXiv:1903.07603 [astro-ph.CO]].

\bibitem{Sherwin:2021vgb}
B.~D.~Sherwin and T.~Namikawa,
[arXiv:2108.09287 [astro-ph.CO]].
	
	
\end{thebibliography}

\end{document}